\documentclass[twocolumn,showpacs,aps,pre,groupedaddress,amssymb,amsmath]{revtex4}

\usepackage{graphicx}
\usepackage{longtable}

\begin{document}

\title{Statistics of microscopic yielding in sheared aqueous foams}

\author{Yuhong Wang}
\author{Kapilanjan Krishan}
\author{Michael Dennin}
\affiliation{Department of Physics and Astronomy, University of
California at Irvine, Irvine, California 92697-4575}

\date{\today}

\begin{abstract}
We detail the statistical distribution of bubble rearrangements in
a sheared two-dimensional foam. Such rearrangements, known as T1
events, are vital to mechanisms resulting in flow through
microscopic mechanical yielding. We find that at a constant rate
of shear, the rate of occurrence of T1 events shows only small
fluctuations. This rate is however seen to vary significantly with
a variation in the initial configuration of bubbles constituting
the foam. In addition, we detail the spatial and orientational
distribution of T1 events and relate them to the distribution of
stresses in the bulk of the material. Some insights into the
irreversibility of the dynamics are also discussed.

\end{abstract}

\maketitle

\section{Introduction}

In many materials, the microscopic structure predominantly
determines the mechanism of flow. The microscopic response to
deformation may vary from linear in the case of elastic materials
and newtonian fluids to highly nonlinear in non-newtonian fluids
and granular materials. In many materials, the response is linear
for sufficiently small deformations, only to become nonlinear at a
critical value of the deformation. The transition to nonlinear
behavior is often controlled by the nature of various microscopic
events. It is worth highlighting just a few examples. In
crystalline solids, defects in the lattice structure may lead to
fractures within the material \cite{gvs01}. In amorphous
materials, plasticity has been described in terms of local
rearrangements of particles, known as shear transformation zones
(STZs) \cite{FL98,ML04}. For soft materials, such as colloids,
foams, and emulsions, a soft glassy rheology (SGR) model based
based on particle rearrangements is able to capture many features
of the macroscopic behavior \cite{SLHC97,S98}. Similarly, in the
context of emulsions, a model based on ``weak'' zones within the
material has been successful in explaining anomalous measurements
of rheological properties \cite{LRMGW96,LL97}. Therefore,
characterizing microscopic rearrangements is imperative to
understanding the behavior of materials subjected to external
loads during stress, strain etc.

\begin{figure}[h!]
\begin{center}
\includegraphics[width=8.5cm]{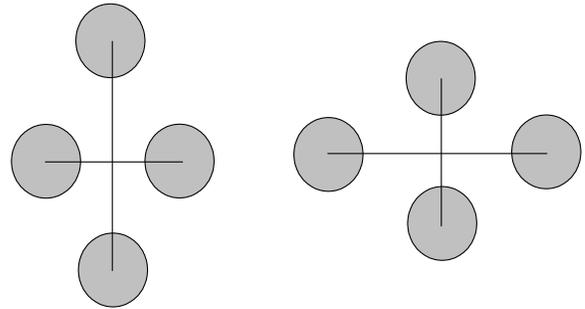}
\end{center}
\caption{The above illustrates the rearrangement dynamics during a
T1 event. The centers of complimentary bubble pairs that undergo
neighbor switching are joined by line segments. The two vertically
connected bubbles become neighbours(N) from being next-nearest
neighbours(NN), symbolized by NN$\rightarrow$N, and vice-versa for
the horizontal bubbles, denoted by NN$\leftarrow$N.}
\label{figureT1}
\end{figure}

In this article we study the behaviour of aqueous foams when
subjected to macroscopic shear. The yielding of such foams is
typically described by a Bingham model that proposes a critical
yield stress at which the material switches from an elastic to a
plastic response \cite{WH99}. When stresses in the material exceed
the yield stress, failure in the microscopic structure results in
flow within the material. The microscopic aspect related to this
behaviour is a succession of slip between the constituent bubbles
of the foam, known as T1 events (see Fig.~\ref{figureT1}). In its
most basic manifestation, such slip involves four neighboring
bubbles, with one pair of nearest neighbours becoming next-nearest
neighbours and vice-versa for another pair. T1 events have been
the subject of study in simulations \cite{WBHA92,OK95,D95,VHC06},
indirect measurement in three-dimensional foam using diffusive
wave spectroscopy (DWS) \cite{DWP91,EJ94,GD95,hch97,chk04}, and
direct observation in two-dimensional foams \cite{DK97,D04,VC05}.

In addition to shear induced T1 events, flowing foams are subject
to a number of other phenomenon such as drainage, aging,
coarsening and T2 events (the disappearance of a bubble within a
foam) \cite{WH99}. While these mechanisms do have an influence on
the flow behaviour, their role is minimal at the short time-scales
of response we are interested in. Various other studies look at
the interaction of these mechanisms with flow, for example
drainage-induced convection \cite{cawh06}.

The role of T1 events resulting in flow in aqueous foams has a
number of interesting implications. The spatial distribution of
these events would provide clues to the distribution of internal
stresses, as T1 events would be expected to occur more frequently
in regions where the stresses exceed the local yield stress of the
material. A similar argument may be used in regard to the
orientation of the T1 events. The slippage between neighbouring
bubbles along a preferred direction indicates an orientational
inhomogeneity within the material or the imposed forces. The
temporal distribution of T1 rates would also be insightful into
relaxation processes that occur: A burst of T1 events between
periods of relative calm could indicate avalanche-type behaviour
wherein stresses are built up slowly over time and released
rapidly \cite{WBHA92,KE99}. And finally, a relationship between
the configuration of bubbles and the spatial and temporal
distribution of T1 events help build connections between the
structural and dynamical properties of such materials.

In investigating these areas, it is important to develop a large
statistics detailing the spatial and temporal distribution of T1
events. Though DWS \cite{DWP91,EJ94,GD95,hch97,chk04} has been
very successful in providing statistical information on individual
bubble motions, there are no well-established techniques to gather
data within the bulk of a three-dimensional foam at the
time-scales and spatial resolution we require to track T1 events
during shear. We therefore focus our attention on a
two-dimensional foam, also known as a bubble raft. The bubble raft
has been found to be a useful guide in the study of foams
\cite{tow00,D04} as well as solid materials \cite{gvs01, sh72,
BL49}. Also, the impact of boundary conditions on T1 events in
bubble rafts has been characterized \cite{VC05}. In addition, it
is helpful to initially study monodisperse bubble rafts. This
makes it easier to interpret the results and eliminates
complexities associated with polydispersity \cite{sh72}. Such
effects may include faster coarsening rates, the development of
spatial inhomogenities due to size dependent flow behaviour of
individual bubbles etc. We also focus our observations in a regime
when the foam is subjected to high rates of shear. It is expected
that the system is not in a quasistatic regime, and the response
is far from equilibrium. In contrast, prior studies have detailed
the response at lower rates of strain \cite{KE99,WH99,D04}, and
the creep response of foam in which the impact of coarsening
induced T1 events is studied \cite{VHC06}.

In the next section we detail techniques of our experiments as
well as image analysis. Following this, we present and discuss our
results on the distribution of T1 events under steady shear.

\section{Experiment and image analysis}

Our experiments utilize a layer of monodisperse bubbles floating
on the surface of water undergoing shear between two
counter-rotating bands. The bubbles were made in a trough
containing an aqueous solution consisting of water, glycerol and
miracle bubble solution in the ratio of 80\%, 15\% and 5\% by
volume respectively. A small amount of fluorescent dye,
Fluorescein, was added as well to enhance imaging within our
system. A steady stream of nitrogen injected into this solution
using a needle was sufficient to generate a monodisperse layer of
bubbles, resulting in a bubble raft.

In addition, two bands with evenly spaced gaps along their length
were positioned at the surface of the bubble monolayer with gears
attached to submerged shafts. The shafts are supported by teflon
bearings attached to  the bottom of the trough. The bands were
attached to a stepper motor through which their rate of motion could
be accurately controlled. The bubbles floating in between these
bands undergo linear shear. The amount of shear is quantified by
$\gamma \ = 2 L(t)/D$, where $D {\rm( = 6cm)}$ is the width between
the bands and $L$ the distance traversed by the band in time $t$.
Typically, the width of the band corresponds to about 20 bubble
diameters. The entire trough was enclosed, and illuminated using UV
lamps. The motion of the bubble raft was recorded as consecutively
saved images using a CCD array. The frame rate of image capture was
maintained high enough so as to identify individual bubbles between
successive images.

The images captured were filtered to remove noise inherent in the
CCD imaging technique as well as some optical inhomogenities
associated with the optics. The filtered images were thresholded to
yield a binary image delineating the location of individual bubbles.
The center positions of individual bubbles were computed as the
center of mass of each of the delineated regions. A
voronoi-construction was then used to construct the edges between
the bubbles constituting the bubble raft. Nearest-neighbours were
designated as those that shared a common edge. Next-nearest
neighbors were also extracted using this information. Individual
bubbles were identified based on the least displacement of bubbles
between consecutive images.

The identification of Nearest-neighbors and next-nearest-neighbors
between successive frames was used to establish the occurrence of
T1 events: i.e. when a set of of nearest neighbors became next
nearest neighbors. The location of the T1 event was designated as
the center of mass of the bubbles participating in the event. A
more detailed account of the experimental and image analysis
techniques may be found in reference \cite{wkd06a}. The
experiments are performed over a number of rates of shear, $\dot
\gamma = d \gamma / \ dt$, producing a sequence of images taken at
equally spaced time intervals.

\section{Results and Discussion}

\subsection{Spatial distribution: Slip zones, homogeneity}
\begin{figure}[htb!]
\begin{center}
\includegraphics[width=8.5cm]{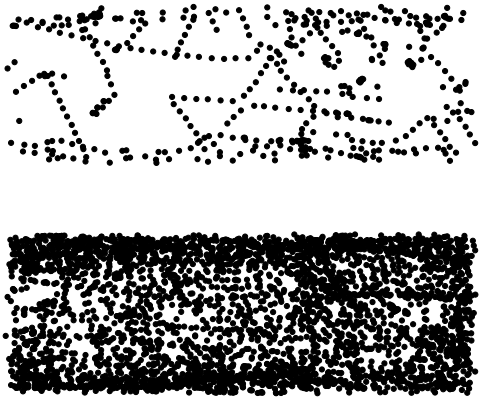}
\end{center}
\caption{The above two plots indicate the positions of T1 events
over different strains. The lower figure is strained about ten
times longer than the upper figure. The top figure indicates the
succession of T1 events that lead to slip. The homogeneous spread
of T1 events is seen in the lower figure. The increase in T1
events close to the driving bands (towards the top and bottom) are
attributed to the pinning of bubbles at these boundaries.}
\label{figureSpread}
\end{figure}

The T1 events are seen to be distributed uniformly between the
bands in the region of interest. There is a slight increase in the
number of T1 events in the vicinity of the bands. This may be
attributed to pinning effects at the bands resulting in locally
higher shear stresses. The spatially uniform distribution of T1
events seen in the lower plot in Fig.~\ref{figureSpread} suggests
that the material properties as well as forces imposed are
homogeneous when averaged over the entire period of shear.

Often, multiple T1 events occurs along a local crystalline axis of
the foam as in the upper plot in Fig.~\ref{figureSpread}. The
spatial extent of such regions may extend across the entire width
of the sheared foam, indicating the role of T1 events in causing
macroscopic slip and flow. Such coherent rearrangements over small
$\gamma$ implies that the local crystalline structure may
influence the focussing of stresses as well as provide principal
directions for fracture over regions of the order of the system
size.

\subsection{Orientational distribution: Anisotropy, irreversibility}

\begin{figure}[htb!]
\begin{center}
\includegraphics[width=8.5 cm]{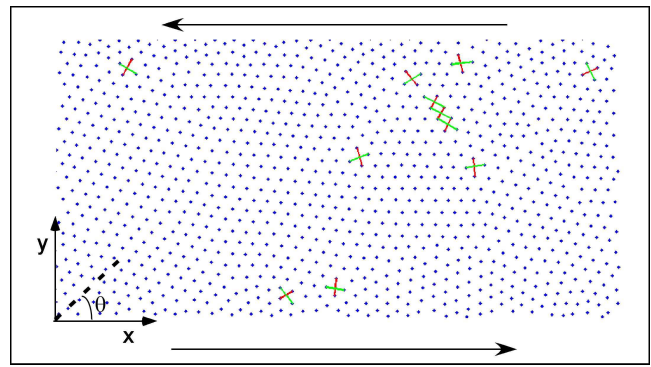}
\end{center}
\caption{The blue dots in the above plot indicate the
instantaneous locations of the centers of bubbles forming the
bubble raft. The red lines connect bubbles that have undergone
compression, becoming nearest neighbors from being next nearest
neighbors (NN$\rightarrow$N). The green lines connect bubbles that
have undergone dilation, becoming next-nearest neighbors from
being nearest neighbors initially (NN$\leftarrow$N). The
orientation of these two events is measured by the angle
($\theta$) the red and blue lines make with the x-axis. The
horizontal arrows at the top and bottom of figure show the
direction of shear imposed by the bands.} \label{figureT1angle}
\end{figure}

\begin{figure}[htb!]
\begin{center}
\includegraphics[width=8.5 cm]{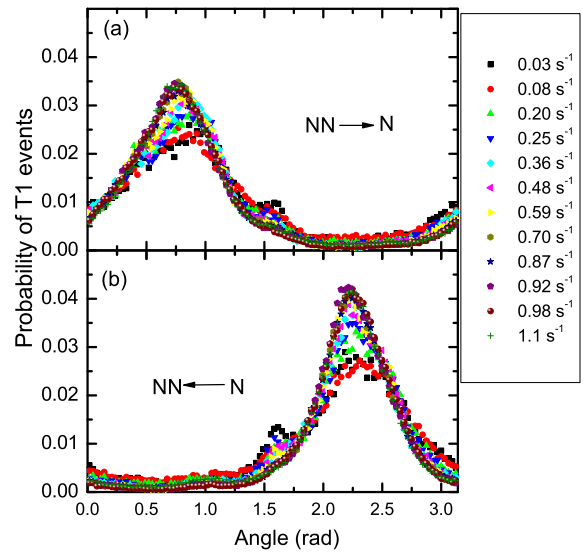}
\end{center}
\caption{The upper figure plots the angle with respect to the
driving bands at which next-nearest bubbles become nearest
bubbles(NN$\rightarrow$N). The lower figure indicates the
complimentary mechanism of NN$\leftarrow$N. The different colours
indicate plots obtained at different rates of shear.}
\label{figureOrientation}
\end{figure}

In order to study the orientation of T1 events, we look at a
single pair of bubbles that participate in the T1 event. This pair
represents the bubbles that become nearest neighbours(N) from
being next-nearest-neighbours(NN).  The geometry used to define
the angles of the T1 events is illustrated in
Fig.~\ref{figureT1angle}. We focus on the angle between the x-axis
(orientation of the driving bands) and the line joining
nearest-neighbors as one parametrization of the orientation of the
T1 events. This is plotted in Fig.~\ref{figureOrientation}a. We
find that the orientation of T1 events shows an distribution
peaked about a particular angle, which may indicate a principal
shear axis. At lower rates of shear, a secondary peak is also
seen. This suggests the existence of a secondary axis for shear
during T1 events.

A similar measurement on the complimentary pairs of bubbles is
shown in Fig.~\ref{figureOrientation}b. The angles corresponding
to the maxima in the curves indicate the most preferred direction
of yielding through T1 events through compression and dilation.
Distinct orientations for bubbles coming closer (NN$\rightarrow$N)
and bubbles moving apart (NN$\leftarrow$N) suggest an anisotropy
in the internal stresses on the bubbles. The difference between
the maxima of angles seen in the upper and lower plots of
Fig.~\ref{figureOrientation} correspond to the angle between the
lines indicated in Fig.~\ref{figureT1}, further illustrating the
role of local structure in directing internal stresses.

In addition to the difference in most probable angular
orientation, there is also a noticeable difference in the widths
of the distributions of Fig.~\ref{figureOrientation}. The upper
plot (NN$\rightarrow$N) is observed to be broader and lower than
the lower (NN$\leftarrow$N). This has important ramifications on
the reversibility of T1 events. In the ideal case of
reversibility, all bubbles undergoing NN$\rightarrow$N would
undergo NN$\leftarrow$N on reversing the shear direction/time. In
such a scenario, the width of angular distributions would remain
unchanged between NN$\rightarrow$N and NN$\leftarrow$N as the two
processes are symmetric under time reversal. Our observation
require that such a reversal would need to occur at different
ranges of angular distributions. This makes the reversibility of
T1 events nontrivial in a statistical sense while placing no
constraints on the reversibility of individual T1 events.

\subsection{Temporal distribution: Rate invariance, extensivity}

\begin{figure}[htb!]
\begin{center}
\includegraphics[width=5.0cm]{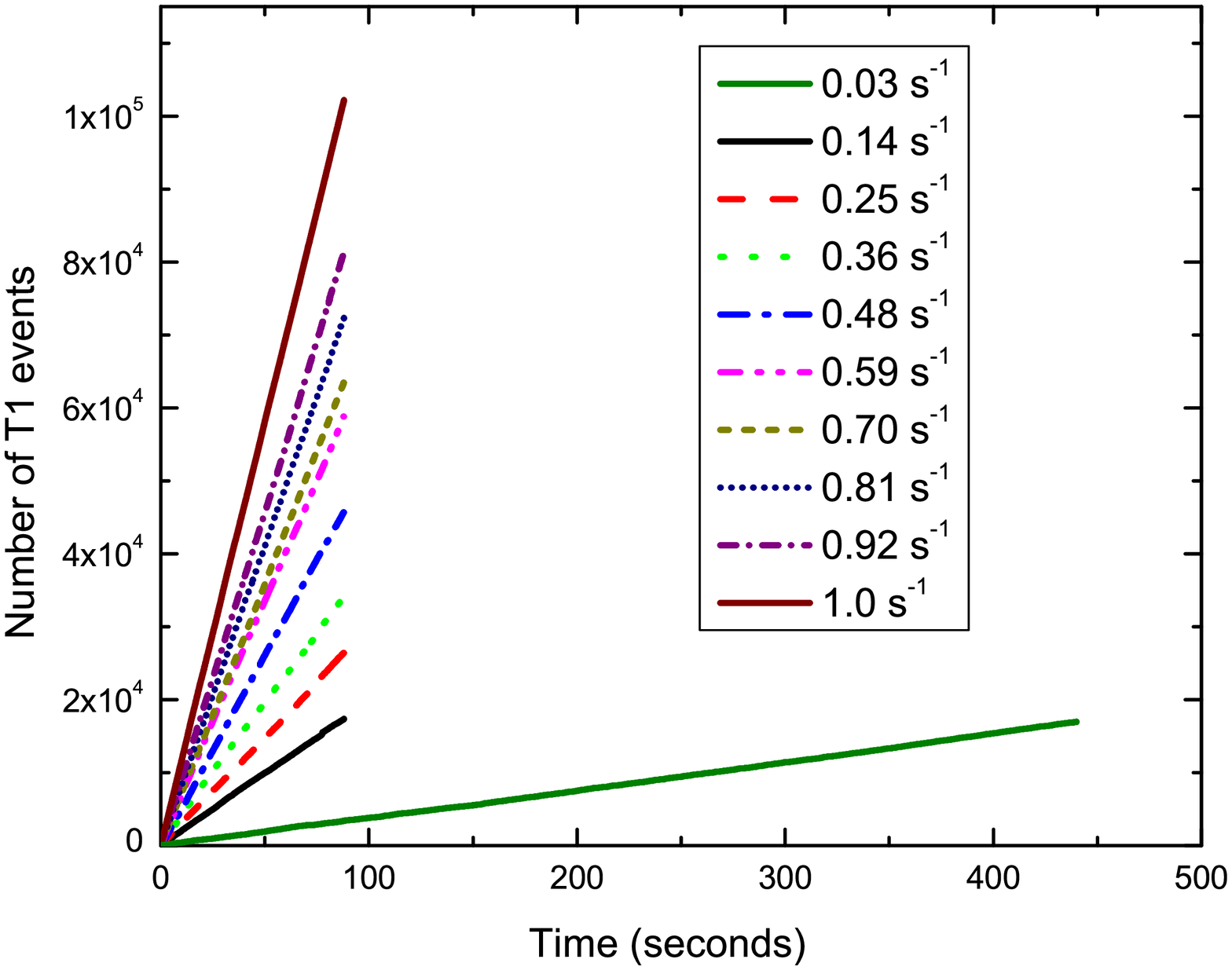}{(a)}
\includegraphics[width=5.0cm]{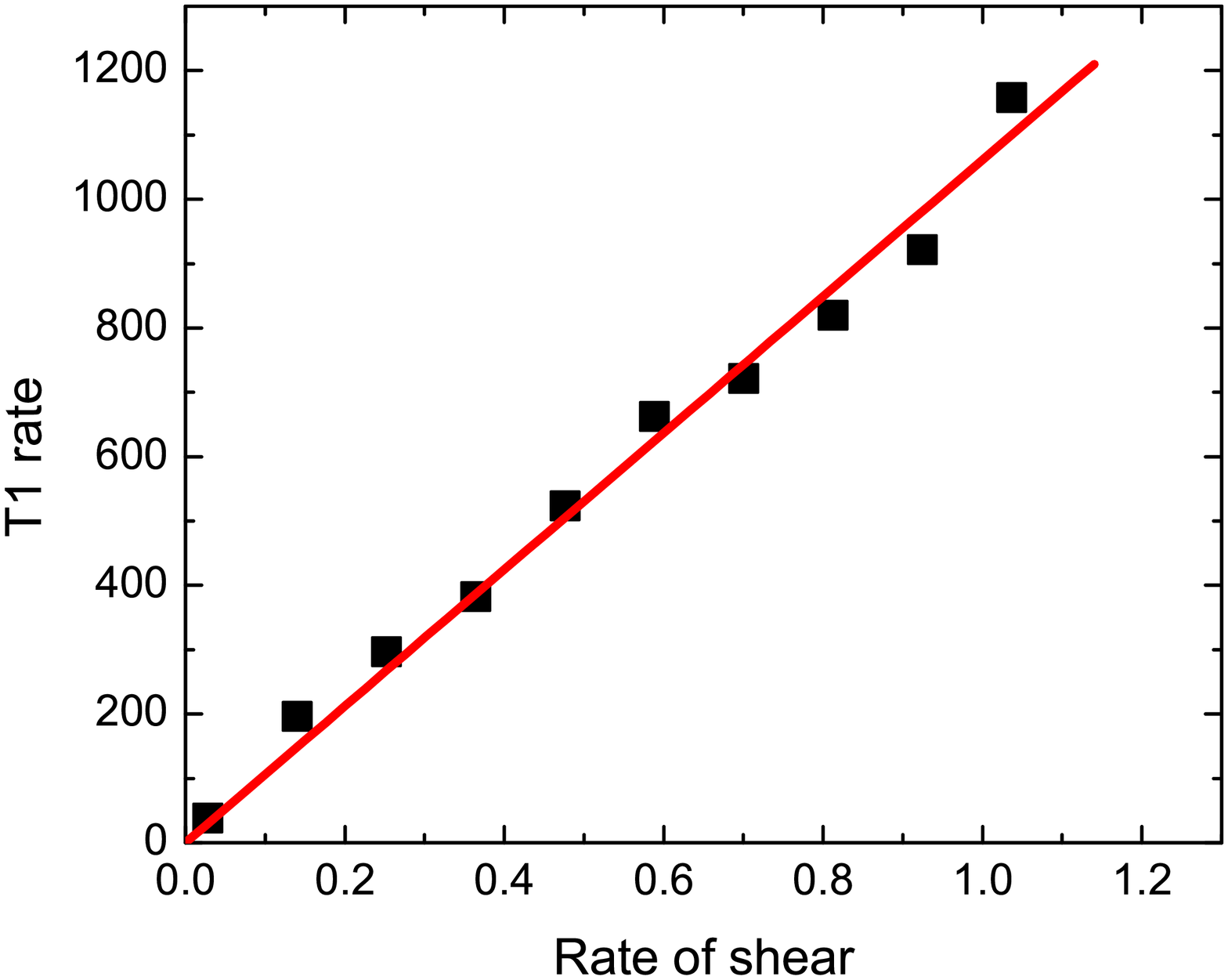}{(b)}
\end{center}
\caption{(a)Number of T1 events vs. time. At the lowest rate of
shear, we also include the data for longer times, illustrating the
rate invariance over longer times. (b)Rate of T1 events vs. rate
of shear. The solid line is a linear fit passing through the
origin.} \label{T1vsTime}
\end{figure}

We also look at the rate at which T1 events occur. We find that
even over large strains, the rate of T1 events is uniform, i.e.
the rate of T1 events does not vary much during the run. It is
particularly noteworthy that the rate of T1 events shows only a
small deviation during shear as indicated in Fig.~\ref{T1vsTime}.
The consistent and well defined slope in upper figure has a number
of implications.

Firstly, it indicates that the rate of T1 events may be considered
equivalent to a short time constraint on the dynamical evolution
of the system. This implies that the response to strain at the
boundaries imposes an internal consistent rate of slippage between
bubbles. The rate of T1 events is seen to increase linearly with
the imposed rate of shear at the boundaries.

Second, when coupled with the homogeneous spatial distribution of
T1 events, it is relevant to talk about the rate of T1 events per
bubble. This implies that the rate of T1 events is an extensive
quantity.

\section{Conclusions and outlook}

We have undertaken a systematic study of T1 events in
two-dimensional sheared aqueous foams. T1 events constitute the
fundamental dynamics associated with foam flows. The study has
emphasized the role of local orientation of bubbles and the
resulting anisotropy in orientational distribution of T1 events.
Future work will focus on connecting the observed dynamics of the
T1 events with the concept of ``weak zones'' developed in the
context of emulsions \cite{LRMGW96,LL97} and the analysis of STZs
in the context of model foams \cite{ML04}. Also, we have
demonstrated a clear correlation between the rate of T1 events and
the rate of strain.

We also open many interesting avenues of investigation. For
example, contrasting the statistics obtained with polydisperse
bubble distributions would be useful in understanding amorphous
materials. Also, the significance of structure at the ends of T1
slip zones in Fig.~\ref{figureSpread} could unravel mechanisms for
fracture nucleation and termination. It would be particularly
interesting to understand the origins of uneven distribution
widths of angular distribution between NN$\rightarrow$N and
NN$\leftarrow$N. Designing a shear geometry that maintains
symmetry in the two distributions would illustrate the influence
of boundaries in determining reversibility. Finally, developing a
better understanding of the connection with the results for
coarsening generated T1 events in a three-dimensional foam
\cite{VHC06} will be an important step.

\begin{acknowledgments}

This work was supported by a Department of Energy grant
DE-FG02-03ED46071.

\end{acknowledgments}


\end{document}